\documentclass{elsart}

\usepackage{natbib}

\usepackage{graphicx}
\usepackage{amssymb}
\begin{document}

\begin{frontmatter}



 \title{Equilibrium relationships for non-equilibrium chemical dependencies}


 \author[parks,langmuir]{G.S.\ Yablonsky\corauthref{cor1}}
 \ead{gyablons@slu.edu}
 \author[cage,dcmoney]{D. Constales}
 \ead{dcons@world.std.com}
 \thanks[dcmoney]{Financial support from BOF/GOA 01GA0405 of Ghent University 
is gratefully acknowledged.}
 \author[lpt]{G.B. Marin}
 \ead{Guy.Marin@UGent.be}
 \corauth[cor1]{Corresponding author.}
 \address[parks]{Parks College, Department of Chemistry, Saint Louis University, 
3450 Lindell Blvd, Saint Louis, MO 63103, USA.}
 \address[langmuir]{The Langmuir Research Institute, 106 Crimson Oaks Court, 
Lake Saint Louis, MO63367, USA}
% \address[wustl]{Dept.\ of Energy, Environmental and Chemical Engineering, 
% Washington
 % University, Campus Box 1198, One Brookings Drive, St. Louis, MO
 % 63130-4899, USA.}
 \address[cage]{Department of Mathematical Analysis, Ghent University, Galglaan 2, 
B-9000 Gent, Belgium.}
 \address[lpt]{Laboratory for Chemical Technology, Ghent University, Krijgslaan 281 
(S5), B-9000 Gent, Belgium.}

 \begin{abstract}
 
In contrast to common opinion, it is shown that equilibrium
constants determine the time-dependent behavior of particular ratios of
concentrations for any system of reversible first-order reactions. Indeed,
some special ratios actually coincide with the equilibrium constant at
any moment in time. This is established for batch reactors, and similar
relations hold for steady-state plug-flow reactors, replacing astronomic
time by residence time. Such relationships can be termed time invariants
of chemical kinetics.

 \end{abstract}


\end{frontmatter}

\def\aru#1#2{\,\stackrel{\mbox{\scriptsize$k^#1_#2$}}{\mbox{\scriptsize$\rightarrow
$}}\,}
\def\ard#1#2{\,\stackrel{\mbox{\scriptsize$\leftarrow$}}{\mbox{\scriptsize$k^#1_#2$}}
\,}

\def\arud#1#2#3#4{\raise-3.5mm\hbox{$\stackrel{\aru#1#2}{\ard#3#4}$}}

\section{Introduction}

In presenting the foundations of physical chemistry, the basic difference
between equilibrium chemical thermodynamics and chemical kinetics is
always stressed.

A typical problem of equilibrium chemical thermodynamics (ECT) is
calculating the composition of a chemical mixture that reacts in a closed
system for an infinitely long time. ECT does not consider time.

In opposite, chemical kinetics is the science of the evolution of chemical
composition in time.

Some selected results of theoretical chemical kinetics are obtained
from thermodynamic principles, especially the principle of detailed
equilibrium:
\begin{itemize}
\item[(a)] the uniqueness and stability of the equilibrium in any closed system,
see \citep{Zeldovich} and the analysis in \citep{Yablonsky1991};

\item[(b)] the absence of damped oscillations near the point of detailed
equilibrium, see \citep{WeiPrater} and the analysis in \citep{Yablonsky1991};

\item[(c)] some limitations on the kinetic relaxation from the given initial
conditions, e.g., based on the known set of equilibrium constants that
determine the equilibrium composition, one can find a forbidden domain
of compositions that is impossible to reach from the given boundary
conditions.
See \citep{Gorban1982,Gorban1984,Gorbanetal2006}.
\end{itemize}

However, the present dogma of physical chemistry holds that it is
impossible to present an expression for any non-steady state chemical
system based on its description under equilibrium conditions, except
for some relations describing the behavior in the linear vicinity of equilibrium.

The goal of this short note is to announce that, contrary to this `dogma',
we have obtained relations of equilibrium type for some non-steady state
chemical systems. This has been achieved for all linear cases (with
a general proof) and some non-linear ones. In a forthcoming extended
paper these relationships will be explained in more detail.

Based on our results, equilibrium thermodynamic relationships can be
considered not only as a description of the final point of temporal
evolution, but as inherent characteristics of the dynamic picture.

The following are examples to illustrate our statement; in all of them we analyze 
traditional models
of chemical kinetics based on the mass-action law. The processes described by 
these models
occur in a closed non-steady-state  chemical system with perfect mixing (batch reactor)
or in an open steady-state chemical system with no radial gradient (plug-flow reactor), whose description is identical to the batch reactor, replacing astronomic time by residence time.

In this short communication, we shall analyze a combination of data of different thought experiments, e.g.
\begin{itemize}
\item a chemical reactor is primed with a substance $A$ only,
\item it is primed with substance $B$ only.
\end{itemize}
In both cases, we shall monitor both concentrations $A$ and $B$ and pay special
attention to the dependencies of ``$B$ produced from $A$'' and ``$A$ produced from $B$''. We shall use the notation $A_A(t)$ for the
temporal concentration dependence of substance $A$, given the initial 
condition $(A,B,\dots)=(1,0,\dots)$, i.e., only $A$ occurs, with normalized 
concentration $1$.
Similarly, $B_A(t)$ is the concentration of substance $B$ for the same initial 
condition;
$A_B(t)$ that of $A$ when at $t=0$, $(A,B,\dots)=(0,1,\dots)$. We found
interesting relationships between these dependencies. 

\section{Linear cases}

\subsection{A single, first-order reversible reaction $A\arud+1-1B$}

By elementary mathematical techniques, for $t\ge0$,
\begin{eqnarray}A_A(t)&=&{k^-_1+k^+_1\exp(-(k^+_1+k^-_1)t)\over k^+_1+k^-_1},\nonumber\\
B_A(t)&=&{k^+_1(1-\exp(-(k^+_1+k^-_1)t))\over k^+_1+k^-_1},\nonumber\\
A_B(t)&=&{k^-_1(1-\exp(-(k^+_1+k^-_1)t))\over k^+_1+k^-_1},\nonumber\\
B_B(t)&=&{k^+_1+k^-_1\exp(-(k^+_1+k^-_1)t)\over k^+_1+k^-_1}.\nonumber\end{eqnarray}

Comparing $B_A$ and $A_B$, it is clear that they are constantly in a fixed 
proportion:
\begin{equation}
{B_A(t)\over A_B(t)}={k^+_1\over k^-_1}=K_{\mbox{\small eq}},
\nonumber\end{equation}
which is the thermodynamic constant of equilibrium. It is remarkable that this ratio 
holds at any moment
of time $t>0$, and not merely in the limit $t\to+\infty$.

\subsection{Two consecutive first-order reactions, the first being reversible: $A\arud
+1-1B\aru+2C$}

Using, e.g., Laplace domain techniques, the following analytical solutions are 
obtained: if we define the expressions
\begin{equation}\lambda_{1,2}={k^+_1+k^-_1+k^+_2\pm\sqrt{(k^+_1+k^-_1+k^
+_2)^2-4k^+_1k^+_2}\over2}\nonumber\end{equation}
and verify that
\begin{equation}\lambda_1>k^+_2>\lambda_2>0,\qquad \lambda_1>k^+_1>
\lambda_2>0, \nonumber\end{equation}
we can write
\begin{eqnarray}
A_A(t)&=&{1\over k^+_2(\lambda_1-\lambda_2)}\left({\lambda_1(k^+_2-\lambda_2)}
\exp(-\lambda_2t)+
{\lambda_2(\lambda_1-k^+_2)}\exp(-\lambda_1t)\right),\nonumber\\
B_A(t)&=&{k^+_1\over\lambda_1-\lambda_2}\left(\exp(-\lambda_2t)-
\exp(-\lambda_1t)\right),\nonumber\\
C_A(t)&=&1-{1\over \lambda_1-\lambda_2}\left({\lambda_1}\exp(-\lambda_2t)-
{\lambda_2}\exp(-\lambda_1t)\right),\nonumber\\
A_B(t)&=&{k^-_1\over\lambda_1-\lambda_2}\left(\exp(-\lambda_2t)-
\exp(-\lambda_1t)\right),\nonumber\\
B_B(t)&=&{1\over k^+_2(\lambda_1-\lambda_2)}\left({\lambda_2(\lambda_1-k^+_2)}
\exp(-\lambda_2t)+
{\lambda_1(k^+_2-\lambda_2)}\exp(-\lambda_1t)\right),\nonumber\\
C_B(t)&=&1-{1\over \lambda_1-\lambda_2}\left({(\lambda_1-k^+_2)}\exp(-
\lambda_2t)+{(k^+_2-\lambda_2)}\exp(-\lambda_1t)\right).
\nonumber\end{eqnarray}
Again, it turns out that $B_A$ and $A_B$ are always in fixed proportion:
\begin{equation}
{B_A(t)\over A_B(t)}={k^+_1\over k^-_1}=K_{\mbox{\small eq}},
\nonumber\end{equation}
and again this is the constant of equilibrium of the $A\leftrightarrow B$ reaction.

\subsection{The cycle of three reversible first-order reactions $A\arud+1-1B\arud
+2-2C\arud+3-3A$}

\begin{figure}[h]
\begin{center}
~~\includegraphics[width=1.0\textwidth]{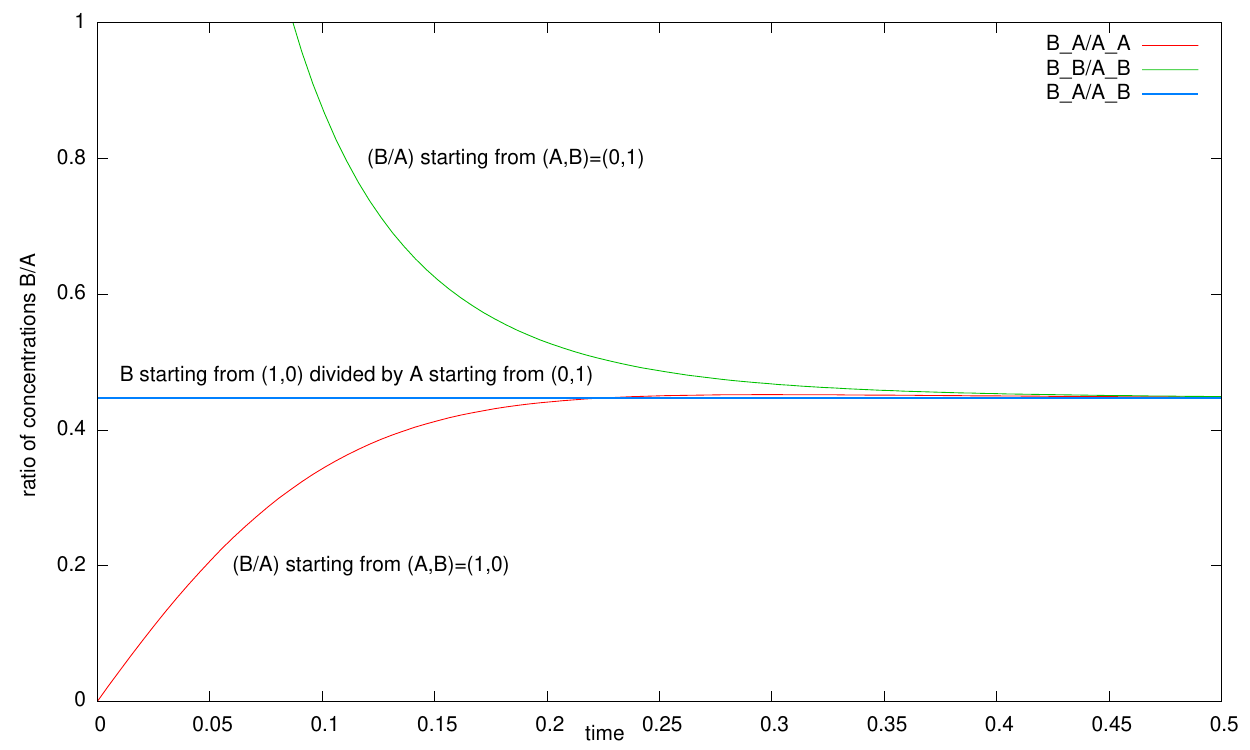}
\end{center}
\caption{\label{fig1}Time dependence of $B_A/A_A$, $B_B/A_B$ and the time-invariant ratio $B_A/A_B$, for the case of isomerization of butenes in 
\citep{WeiPrater}, eq.\ (129).}
\end{figure}

In the Laplace domain, defining the symbols
\begin{eqnarray}
\sigma_1&=&k^+_1+k^-_1+k^+_2+k^-_2+k^+_3+k^-_3\nonumber\\
\sigma_2&=&k^+_1k^+_2+k^+_2k^+_3+k^+_3k^+_1+k^+_1k^-_2+k^+_2k^-_3+k^
+_3k^-_1
+\nonumber\nonumber\\&&k^-_1k^-_3+k^-_2k^-_1+k^-_3k^-_2\nonumber\\
\Delta(s)&=&s(s^2+\sigma_1s+\sigma_2),
\nonumber\end{eqnarray}
the transformed concentrations $A_A$, $B_A$ and $A_B$ are given by
\begin{eqnarray}
{\cal L}A_A(s)&=&{s^2+(k^-_1+k^+_3+k^+_2+k^-_2)s
+(k^-_1k^+_3+k^-_1k^-_2+k^+_2k^+_3)\over \Delta(s)}\nonumber\\
{\cal L}B_A(s)&=&{k^+_1s+(k^+_1k^+_3+k^+_1k^-_2+k^-_2k^-_3)
\over\Delta(s)}\nonumber\\
{\cal L}A_B(s)&=&{k^-_1s+(k^-_1k^+_3+k^-_1k^-_2+k^+_2k^+_3)
\over\Delta(s)}
\nonumber\end{eqnarray}
and the ratio of the latter two, by
\begin{eqnarray}
{{\cal L}B_A(s)\over{\cal L}A_B(s)}&=&
{k^+_1s+(k^+_1k^+_3+k^+_1k^-_2+k^-_2k^-_3)\over
k^-_1s+(k^-_1k^+_3+k^-_1k^-_2+k^+_2k^+_3)}\nonumber\\
&=&{k^+_1\over k^-_1}\left(1-{1\over\displaystyle1+k^-_1{k^+_1k^+_3+k^+_1k^-
_2+k^-_2k^-_3+sk^+_1
\over k^+_1k^+_2k^+_3- k^-_1k^-_2k^-_3}}\right)\nonumber\\
&=&{k^+_1\over k^-_1},
\nonumber\end{eqnarray}
where the Onsager relationship $k^+_1k^+_2k^+_3=k^-_1k^-_2k^-_3$ was used in 
the final step. Hence for
all $s$, ${\cal L}B_A(s)$ and ${\cal L}A_B(s)$ are in fixed proportion given by the 
equilibrium constant $(k^+_1/ k^-_1)$. Since the inverse Laplace transform is linear, 
the same proportion holds in the time domain:
\begin{equation}
{B_A(t)\over A_B(t)}={k^+_1\over k^-_1}=K_{\mbox{\small eq,1}},
\nonumber\end{equation}
Similarly,
\begin{equation}
{C_B(t)\over B_C(t)}={k^+_2\over k^-_2}=K_{\mbox{\small eq,2}},
\quad
{A_C(t)\over C_A(t)}={k^+_3\over k^-_3}=K_{\mbox{\small eq,3}},
\nonumber\end{equation}
As an example, we show in Fig.\ \ref{fig1} the time dependence of $B_A/A_A$, 
$B_B/A_B$ and the time-invariant ratio $B_A/A_B$, for the case of isomerization of 
butenes in \citep{WeiPrater}, eq.\ (129): 
$$
\mbox{{\em cis}-2-butene}
\raise-3.5mm\hbox{$\stackrel{\,\stackrel{\mbox{\scriptsize$4.623$}}{\mbox{\scriptsize$\rightarrow
$}}\,}{\,\stackrel{\mbox{\scriptsize$\leftarrow$}}{\mbox{\scriptsize$10.344$}}
\,}$}
\mbox{1-butene}
\raise-3.5mm\hbox{$\stackrel{\,\stackrel{\mbox{\scriptsize$3.724$}}{\mbox{\scriptsize$\rightarrow
$}}\,}{\,\stackrel{\mbox{\scriptsize$\leftarrow$}}{\mbox{\scriptsize$1.000$}}
\,}$}
\mbox{{\em trans}-2-butene}
\raise-3.5mm\hbox{$\stackrel{\,\stackrel{\mbox{\scriptsize$3.371$}}{\mbox{\scriptsize$\rightarrow
$}}\,}{\,\stackrel{\mbox{\scriptsize$\leftarrow$}}{\mbox{\scriptsize$5.616$}}
\,}$}
\mbox{{\em cis}-2-butene}.
$$

This behavior is typical of the examples 
given here: if the system is started from the initial values $(A,B,C)=(1,0,0)$, 
the ratio  $B_A/A_A$  grows  at first from $0$, eventually reaching the equilibrium 
value (note that for these parameters $B_A/A_A$ slightly overshoots the limit). 
Similarly, when started from $(0,1,0)$, the corresponding $B_B/A_B$
initially decreases from $+\infty$ and eventually reaches
same limit, viz the equilibrium ratio.  But surprisingly the combination $B_A/A_B$ is 
constantly equal to the equilibrium value, for all times $t>0$.

\subsection{The cycle of four reversible first-order reactions $A\arud+1-1B\arud
+2-2C\arud+3-3D\arud+4-4A$}

Although the expressions become more involved, it is straightforward to verify that in 
this case the fixed
equilibrium proportions also hold, given the Onsager relationship
$k^+_1k^+_2k^+_3k^+_4=k^-_1k^-_2k^-_3k^-_4$:
\begin{equation}
{B_A(t)\over A_B(t)}={k^+_1\over k^-_1}=K_{\mbox{\small eq,1}},
\nonumber\end{equation}
and similarly for $C_B/B_C$, $D_C/C_D$ and $A_D/D_A$. Furthermore
\begin{equation}
{C_A(t)\over A_C(t)}={k^+_1k^+_2\over k^-_1k^-_2}=
K_{\mbox{\small eq,1}}K_{\mbox{\small eq,2}},
\nonumber\end{equation}
with a similar relationship for $D_B/B_D$.

\subsection{Sketch of proof for general systems of first-order reactions}

We use the terminology and results of \citep{Roelantetal2010}. Let $A$ and $B$ be 
substances such that a path from either to the other exists. In view of Onsager 
relations, then there must exist a reversible path from $A$ to $B$. The denominators 
of $A_B$ and $B_A$ in the Laplace domain are the same, but their numerators 
differ. For $A_B$, the numerator consists of contributions from all forests where $A$ 
is the root of a tree and $B$ is in that tree. Let $Z$, $X$, $Y$ denote other nodes as 
in the denominator of the left-hand term in (\ref{xform}):
\begin{equation}
{Z\rightarrow A \rightarrow X \rightarrow B \leftarrow Y\over
Z\rightarrow A \leftarrow X \leftarrow B \leftarrow Y}={k_{A \rightarrow X}k_{X 
\rightarrow B}\over k_{X\rightarrow A}k_{B \rightarrow X}}=K_{A\rightarrow\dots
\rightarrow B}
\label{xform}
\nonumber\end{equation}
Reversing the $X\to A$ and $B\to X$ arrows amounts to multiplying the forest's term by $
{(k_{A \rightarrow X}k_{X \rightarrow B})/(k_{X\rightarrow A}k_{B \rightarrow X})}$. In 
view of the Onsager relations, this value does not depend on $X$, but only on $A$ and 
$B$; in fact, it is the equilibrium constant $K_{A\rightarrow\dots\rightarrow B}$. If several $X$ or 
$Y$ or $Z$ occur, the reasoning is the same. Consequently, every term in the 
numerator of $A_B$ is in that proportion to the corresponding term in the numerator 
of $B_A$, and the fixed proportion for all $s$ translates directly to the time domain:
\begin{equation}
{B_A(t)\over A_B(t)}=K_{A\rightarrow\dots\rightarrow B}.
\nonumber\end{equation}

\section{Nonlinear cases}

\subsection{Nonlinear reversible reaction (forward second order, backward first 
order) $2A\arud+1-1B$}

In accordance to the mass conservation law, the balance is
\begin{eqnarray}
A(t)+2B(t)&=&1
\nonumber\end{eqnarray}
In order for the $A$ and $B$ trajectories to reach the same equilibrium, we choose 
to start them from $(1,0)$ and $(0,1/2)$ respectively. The nonlinear differential 
equation is
\begin{eqnarray}
{dA(t)\over dt}&=&-2k^+_1A^2(t)+k^-_1(1-A(t))
\nonumber\end{eqnarray}
which can be solved analytically as
\begin{eqnarray}
A_A(t)&=&{\displaystyle\sqrt{8{k^+_1\over k^-_1}+1}+\tanh\left(
{1\over2}tk^+_1\sqrt{8{k^+_1\over k^-_1}+1}\right)
\over\displaystyle\sqrt{8{k^+_1\over k^-_1}+1}+
\left(4{k^+_1\over k^-_1}+1\right)\tanh\left(
{1\over2}tk^+_1\sqrt{8{k^+_1\over k^-_1}+1}\right)}\nonumber\\
B_A(t)&=&{\displaystyle2{k^+_1\over k^-_1}\tanh\left(
{1\over2}tk^+_1\sqrt{8{k^+_1\over k^-_1}+1}\right)
\over\displaystyle\sqrt{8{k^+_1\over k^-_1}+1}+
\left(4{k^+_1\over k^-_1}+1\right)\tanh\left(
{1\over2}tk^+_1\sqrt{8{k^+_1\over k^-_1}+1}\right)}\nonumber\\
A_B(t)&=&{\displaystyle2\tanh\left(
{1\over2}tk^+_1\sqrt{8{k^+_1\over k^-_1}+1}\right)
\over\displaystyle\sqrt{8{k^+_1\over k^-_1}+1}+\tanh\left(
{1\over2}tk^+_1\sqrt{8{k^+_1\over k^-_1}+1}\right)}.
\nonumber\end{eqnarray}
The remarkable proportion in this case differs slightly from the linear examples:
\begin{eqnarray}
{B_A\over A_AA_B}&=&{k^+_1\over k^-_1}=K_{\mbox{\small eq,1}}
\nonumber\end{eqnarray}
we see that the denominator involves the $A$ concentrations of both trajectories, 
$A_A$ and $A_B$. Only this
can ensure a ratio that equals the equilibrium constant at every time $t>0$.

\subsection{Nonlinear reversible reaction (forward and backward second order) 
$2A\arud+1-12B$}

The mass conservation law is
\begin{eqnarray}
A(t)+B(t)&=&1,
\nonumber\end{eqnarray}
which offers no difficulties for the initial values, $(1,0)$ and $(0,1)$. The differential 
equation
\begin{eqnarray}
{dA(t)\over dt}&=&-2k^+_1A^2(t)+2k^-_1(1-A(t))^2
\nonumber\end{eqnarray}
can be solved analytically as
\begin{eqnarray}
A_A(t)&=&{1\over\displaystyle1+\sqrt{k^+_1\over k^-_1}\tanh\left(2t\sqrt{k^+_1k^-
_1}\right)}\nonumber\\
B_A(t)&=&{\displaystyle\sqrt{k^+_1\over k^-_1}\tanh\left(2t\sqrt{k^+_1k^-_1}\right)
\over\displaystyle1+\sqrt{k^+_1\over k^-_1}\tanh\left(2t\sqrt{k^+_1k^-_1}\right)}\nonumber\\
A_B(t)&=&{\displaystyle\sqrt{k^-_1\over k^+_1}\tanh\left(2t\sqrt{k^+_1k^-_1}\right)
\over\displaystyle1+\sqrt{k^-_1\over k^+_1}\tanh\left(2t\sqrt{k^+_1k^-_1}\right)}\nonumber\\
B_B(t)&=&{1\over\displaystyle1+\sqrt{k^-_1\over k^+_1}\tanh\left(2t\sqrt{k^+_1k^-
_1}\right)}.
\nonumber\end{eqnarray}
Again eliminating time, the similar proportion for this case is
\begin{eqnarray}
{B_AB_B\over A_AA_B}&=&{k^+_1\over k^-_1}=K_{\mbox{\small eq,1}},
\nonumber\end{eqnarray}
where both numerator and denominator have undergone a duplication in $A$ and 
$B$ trajectories, still producing
the equilibrium constant at all times $t>0$.

\section{Conclusions}

We are going to describe these results in detail in a full-length paper.
Then the similar approach will be applied to different systems, i.e.,
CSTR reactors, to reaction-diffusion TAP systems, etc.

Presenting simply the result of this paper, it is a surprising relationship between $A$ from $B$ and $B$ from $A$. An ``ABBA rule''  for short, which reminds the authors of their youth.

\begin{center}
~\hfill {\em ``They speak strangely but I understand'' --- ABBA, "Eagle" (1978).}
  \end{center}

\newpage

\subsubsection*{Caption to figure}

\end{document}